\documentclass{ar-1col-S2O}
\usepackage[utf8]{inputenc}
\usepackage{textgreek}
\DeclareUnicodeCharacter{03BC}{\textmu}
\usepackage[numbers]{natbib}
\usepackage{amsmath,amssymb,amsfonts}%
\usepackage{url}
\jname{Xxxx. Xxx. Xxx. Xxx.}
\jvol{AA}
\jyear{YYYY}
\doi{10.1146/((please add article doi))}

\begin{document}
%TC:ignore
% Page header
\markboth{Almani et al.}{Embodied sensorimotor control}

% Title
\title{Embodied sensorimotor control: computational modeling of the neural control of movement}

%Authors, affiliations address.
\author{Muhammad Noman Almani,$^{1,2,4*}$ John Lazzari,$^{1,3,5*}$ Jeff Walker,$^{1,3,6*}$ and Shreya Saxena$^{1,2,3,7}$
\affil{$^1$Center for Neurocomputation and Machine Intelligence, Wu Tsai Institute, Yale University, New Haven, USA, 06511}
\affil{$^2$Department of Electrical and Computer Engineering, Yale University, New Haven, USA, 06511}
\affil{$^3$Department of Biomedical Engineering, Yale University, New Haven, USA, 06511}
\affil{$^4$email address: muhammadnoman.almani@yale.edu}
\affil{$^5$email address: john.lazzari@yale.edu}
\affil{$^6$email address: jeffrey.walker@yale.edu}
\affil{$^7$email address: shreya.saxena@yale.edu}
\affil{$^*$These authors contributed equally.}}

%Abstract

\begin{abstract}
We review how sensorimotor control is dictated by interacting neural populations, optimal feedback mechanisms, and the biomechanics of bodies. First, we outline the distributed anatomical loops that shuttle sensorimotor signals between cortex, subcortical regions, and spinal cord. We then summarize evidence that neural population activity occupies low-dimensional, dynamically evolving manifolds during planning and execution of movements. Next, we summarize literature explaining motor behavior through the lens of optimal control theory, which clarifies the role of internal models and feedback during motor control. Finally, recent studies on embodied sensorimotor control address gaps within each framework by aiming to elucidate neural population activity through the explicit control of musculoskeletal dynamics. We close by discussing open problems and opportunities: multi-tasking and cognitively rich behavior, multi-regional circuit models, and the level of anatomical detail needed in body and network models. Together, this review and recent advances point towards reaching an integrative account of the neural control of movement.
\end{abstract}

%Keywords, etc.
\maketitle

%Table of Contents
\tableofcontents

%TC:endignore
% Heading 1
\section{INTRODUCTION}
How do distributed neural circuits drive purposeful movements from the complex musculoskeletal system? This characterization is critical towards not just furthering our understanding of the generation of movement, but, importantly, guiding us towards therapeutic targets for diseases affecting motor control. The neural processes leading to movements have been relatively well posited and understood due to the quantitative nature of the behavioral outputs involved. Classic approaches have largely focused on optimization principles, including limb control, to achieve human-like behavioral trajectories. These largely theoretical models of sensorimotor control can recapitulate observed movement trajectories by hypothesizing the presence of a controller guiding the complex movements. However, these models cannot predict how neuronal populations in each brain region affects the resulting movement and vice-versa. On the other hand, breakneck advances in hardware techniques have led to vast improvements in our ability to record large-scale multi-regional neural data. These recordings have enabled dimensionality reduction and modeling techniques to elucidate the structure in high-dimensional neural activity during different conditions, and relate the neural activity directly to kinematic outcomes. However, these data-driven models typically do not consider the biophysical underpinnings of the musculoskeletal system, and thus fail to elucidate the computational role of neural activity in driving the musculoskeletal system such that the body reaches a desired state. The emerging field of embodied control incorporating detailed musculoskeletal models and integrating them with neural computations aim to address these gaps.

In this review, we (1) outline the anatomical substrates of sensorimotor control; (2) examine how neural dynamics are quantified in motor regions; (3) map the theoretical framework of optimal control onto sensorimotor processing; and (4) survey recent and ongoing efforts to explicitly incorporate embodiment into models of sensorimotor control. We end with directions of future work and open questions in the field.

\section{RELEVANT ANATOMY OF THE SENSORIMOTOR CONTROL LOOP}
\begin{figure}[h]
\includegraphics[width=0.7\textwidth]{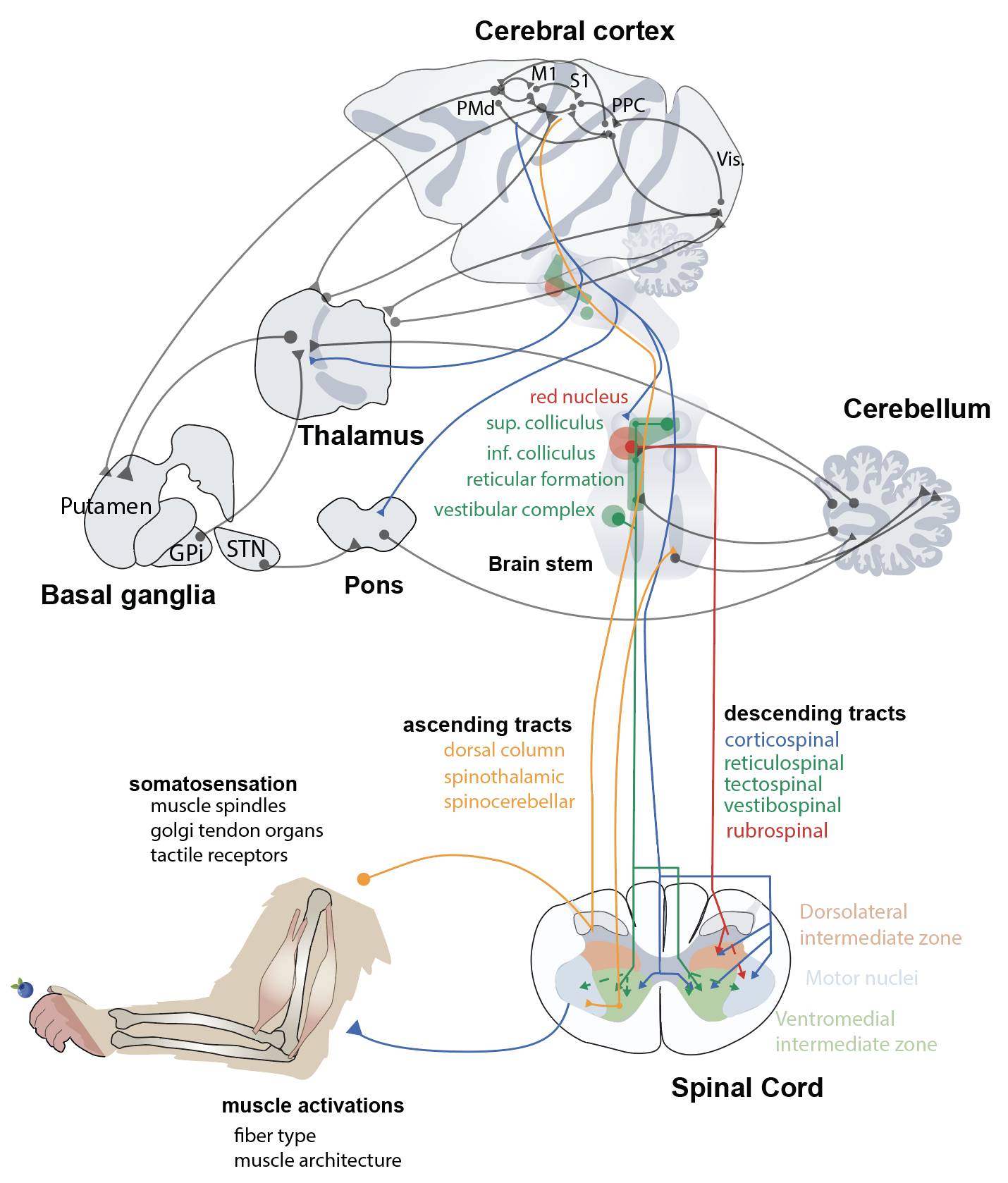}
\caption{{\bf Schema of structures and pathways involved in sensorimotor control of upper limb behavior in the macaque monkey.}  Descending and ascending tracts comprise central conduit for sensorimotor signal flow. Long range projections and slow muscles induce sensorimotor delays. Note multiple examples of anatomical features thought to create copies of motor commands that could provide elements needed to learn predictive internal models for sensorimotor control: corticospinal collaterals to subcortical structures \cite{sherman_thalamus_2016, arber_connecting_2018}, V2b interneurons and spinocerebellar loops \cite{ruder_brainstem_2019}.  Also note nested loops through distributed circuits: intracortical frontoparietal, corticothalamic, cortico-basal ganglia-thalamo-cortical, spinocerebellar, basal ganglia-cerebellar inter thalamic \cite{bostan_basal_2018}. PMd: dorsal premotor cortex, M1: primary motor cortex, S1: primary sensory cortex, PPC: posterior parietal cortex, GPi: globus paladus internus, STN: subthalamic nucleus, Vis.: visual cortical areas. Adapted from \cite{ lemon_descending_2008, ruder_brainstem_2019, sherman_thalamus_2016, battaglia-mayer_corticocortical_2019, bostan_basal_2018}.}
\label{fig1}
\end{figure}

The abstraction of a sensorimotor loop is often used to model the parallel and distributed neuromechanical systems that come together to implement sensorimotor control in humans and other animals.  Although the concept of a single sensorimotor loop does not naturally simplify the complexities of hierarchical and nested biological neural circuits, it does provide a useful framework for modeling goal-directed movements of bodies in the world. Below we will briefly describe components of primate motor systems to illustrate the neural structures involved in biological sensorimotor control and the ubiquity of sensorimotor loops connecting them (\textbf{Figure \ref{fig1}}). The central loop comprises ascending pathways of the spinal cord that carry somatosensory signals from the periphery and descending pathways that carry motor signals to control the musculature.  These pathways are embedded in hierarchically nested cortical, subcortical, and spinal structures that modulate and use those signals for various learning processes to support flexible behavior.  Sensorimotor delays resulting from relatively slow muscle activation and long-range signal transduction create the need for predictive mechanisms to compensate.  We find the ingredients needed to build such mechanisms across both cortical and subcortical structures.

\subsection{Sensory Regions}
\subsubsection{Ascending projections: sources and targets}
Sensory feedback is an essential part of naturalistic motor control.  Afferent feedback resulting from movement of the body or world propagates through a sequence of circuits originating in peripheral receptors, traveling through the dorsal root ganglion and ascending the spinal cord and brainstem toward somatotopically structured thalamocortical loops for visual and somatosensory processing.  The most immediately relevant areas for discussions of sensorimotor control are the cortical areas, 3b, 3a and area 2, spanning anterior and posterior parietal cortex.  These areas are involved in processing tactile input from receptors in glabrous skin (mainly primary sensory area 3b) and proprioceptive muscle spindle and golgi tendon organ inputs from within muscles and tendons (primarily 3a and area 2) ascending the dorsal column of the spinal cord, entering the brain through the cuneate nucleus before passing through thalamic nuclei on their way to the cerebral cortex \cite{delhaye_neural_2018, versteeg_cuneate_2021}. Population analysis of responses in these cortical areas has revealed that these neurons encode detailed information about kinematic limb state (e.g. \cite{chowdhury_area_2020, goodman_postural_2019}). Although the exact nature of the connection between S1 and the motor regions remains a topic of ongoing study, anatomical work demonstrates strong reciprocal connections, both intracortical and transthalamic \cite{gomez_parallel_2021}. 

\subsubsection{Vision and posterior parietal cortices}
Various brain regions involved in visual processing and visuomotor coordinate transformations are critical for visually guided motor control involving complex interactions with the external environment. For example, areas within the posterior parietal cortex (PPC) receive a significant amount of visual information, share extensive connections with premotor areas, and have been shown to be crucial for reaching to visual targets in humans and nonhuman primates \cite{battaglia-mayer_corticocortical_2019, bufacchi_cortico-spinal_2023}.    

\subsection{Motor Regions}
\subsubsection{Descending projections: sources and targets}
Parallel and distributed circuits spanning cortical and subcortical structures underpin the sophisticated control of movement observed in both humans and other primates.  We can make sense of these circuits in terms of the composition of descending projections of the spinal cord \cite{lemon_descending_2008}.  Reticulospinal, vestibulospinal, and tectospinal fibers originate from the brain stem and terminate largely in the intermediate zone of the spinal cord. Rubrospinal fibers originate in the red nucleus and terminate in the ventromedial intermediate zone as well as in the motor pools of the spinal cord. These pathways are strongly conserved, provide an evolutionary foundation for the control of voluntary movement, provide redundancy in spinal access that can compensate for injury, and are critical for any comprehensive understanding of vertebrate motor control \cite{moreno-lopez_sensorimotor_2016}.  We know that in mice, these circuits are spatially organized to allow access to components of movement, for example, a reach or a grasp \cite{yang_structural_2023}.  However, while the spatial organization and function of many brain stem circuits have been mapped in exquisite detail in mice \cite{arber_connecting_2018, ruder_brainstem_2019}, their population dynamics in primates has been largely unexplored.  In contrast, there has been an explosion of interest in dynamics of neural populations from which the corticospinal tract originates, specifically primary motor cortex, premotor cortex, but also primary sensory areas just posterior to the central sulcus, as well as cingulate motor areas and the supplementary motor area.  Neural populations in these cortical areas will be the focus of much of the discussion about population dynamics in \textbf{Section \ref{ch:neur_dyn}}. Corticospinal projections from these areas terminate in both the intermediate zone and the motor pools of the spinal cord. These populations are potentially 1-2 synapse from neuromuscular junctions and are known to be critical for dexterous and visually guided voluntary movement \cite{lemon_descending_2008}.

\subsubsection{Cerebellum, Basal ganglia and Thalamus}
Although they do not project directly onto the spinal cord, the basal ganglia and cerebellum are critical to sensorimotor function. The cortical-basal ganglia (BG)-thalamocortical (CBGTC) loop contains multiple direct and indirect feedback loops implicated in a wide array of motor functions~\cite{mink1996basal}. The frontal cortex projects densely to the striatum, and striatal efferents converge at the BG output nuclei such as the Substantia Nigra reticulata or Globus Pallidus internus. Output nuclei inhibition (excitation) then disinhibits (inhibits) the thalamus which subsequently excites (inhibits) the cortex. In addition to classical corticostriatal projections known to be involved in task selection, motor control, and learning, recent work using the rabies virus to trace multisynaptic pathways from the motor cortex, cerebellum, and BG has revealed a reciprocal basal ganglia cerebellar loop of projections primarily through the ventrolateral thalamic nucleus and the pontine nucleus of the pons \cite{bostan_basal_2018}. These subcortical structures contain many anatomical features that suggest elements of feedback control of the sort discussed in \textbf{Section  \ref{ch:opt_ctrl}}. Corticospinal projections are known to provide collaterals to subcortical structures on their way to their spinal targets, such as the pontine nucleus of the pons; these are thought to carry copies of motor commands \cite{arber_connecting_2018}. Similarly, V2b interneurons in ventromedial intermediate zone of the spinal cord bifurcate on their way to the motor pools and ascend the spinal cord to provide a copy of this motor command to the lateral reticular nucleus \cite{ruder_brainstem_2019}.  In both of these cases, the structures that receive the motor command copy subsequently project to different components of the cerebellum and deep cerebellar nuclei.

\subsection{Musculoskeletal System}
All signals bound to activate muscles must go through the motor pools of the spinal cord, the final common path. These motor pools are somatotopically organized and engage alpha motor neurons to activate motor units and muscle fibers \cite{taitano_muscle_2024}.  The muscle fibers are organized into three-dimensional joint-spanning muscle bodies in series with tendons such that force production for a given set of muscle activations will be a function of the bony origins and attachments of the muscles, the arrangement and composition of muscle fibers in the muscle bodies, and the current state of the joint \cite{zajac_muscle_1989}.  Muscle spindles are embedded in parallel with muscle fibers to sense passive and active joint stretch.  The sensitivity of these muscle spindles is itself regulated by a system of gamma motor neurons \cite{delhaye_neural_2018}.  Golgi tendon organs are embedded in tendinous musculoskeletal attachments and work in concert with inhibitory interneurons in the spinal cord to sense and protect against forceful muscle contraction.  While the description of muscle physiology is necessarily brief given the scope of this review, for a more comprehensive review of muscle physiology especially modeling at the neuromuscular interface see \cite{rohrle_multiscale_2019}.

Circuits and structures underlying sensorimotor control in primates and other vertebrates comprise a parallel and distributed neuromechanical system where each component contains sensorimotor loops and anatomical features consistent with feedback control processes being distributed throughout.  This presentation of relevant anatomy drew primarily from work with primates, but also from work with rodents for cellular and circuit level characterizations.  We now turn our attention to an understanding of the population dynamics that has largely emerged from studies of the activity of motor cortical populations.

% Neural population dynamics
\section{NEURAL POPULATION DYNAMICS}
\label{ch:neur_dyn}
Here we discuss the hypothesis that movement is generated by an underlying high-dimensional, non-linear dynamical system implemented by populations of neurons. This theory of neural computation now permeates systems neuroscience, guiding theories of motor and associated cognitive processes such as memory, decision-making, and preparation.

\begin{figure}[h]
\includegraphics[width=0.95\textwidth]{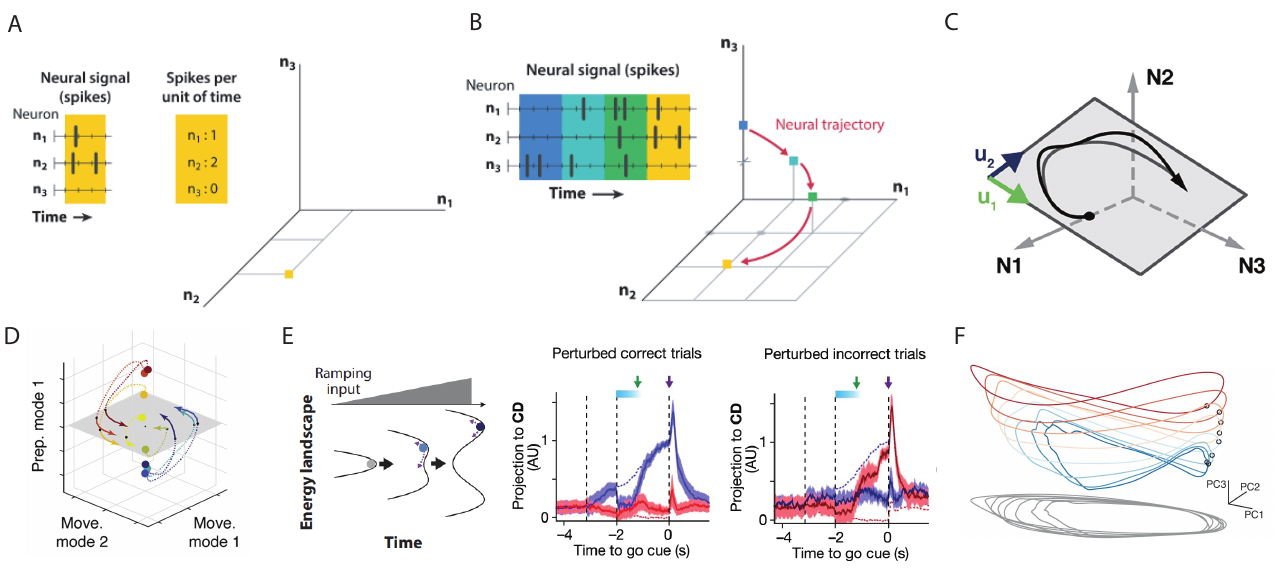}
\caption{{\bf Computation through neural population dynamics.} (A) The firing rate of each neuron in the population defines a (high dimensional) neural state space. Adapted from \cite{2020vyasComputationNeuralPopulation}. (B) As neural firing changes over time, an \(N\) dimensional curve is traced through the state space. Adapted from \cite{2020vyasComputationNeuralPopulation}. (C) The neural trajectory generated during behavior may lie on a lower dimensional manifold within the \(N\) dimensional space. In this case, the manifold is linear and two-dimensional, defined by the modes \(u_1\) and \(u_2\). Adapted from \cite{gallego2017neural}. (D) Neural trajectories during complex behavior may shift between subspaces. The latent activity along the first preparatory mode (dotted line) is nearly orthogonal to the latent activity defining the movement space (filled line). Adapted from \cite{elsayed2016reorganization}. (E) Optogenetic perturbations of mouse ALM activity during motor preparation either recovers, or switches its position along the choice axis after stimulation. (left) Schematic of externally driven discrete attractor guided by a ramping signal. (middle) Recovery of perturbed activity along a choice axis during correct lick trials. (right) Perturbation trials resulting in incorrect choices show activity flipping along the choice axis. Adapted from \cite{inagaki2019discrete}. (F) During cyclic movements of the primate limb, activity in M1 displays elliptical dynamics stacked according to the speed of movement. Adapted from \cite{ 2022saxenaMotorCortexActivity}.}
\label{fig2}
\end{figure}

% Computation through dynamics
\subsection{Computation Through Dynamics}

The computations performed by the motor cortex used to produce volitional movements have traditionally been studied using a bottom-up approach. In this framework, the tuning of single-unit responses to external variables such as muscle activity or movement kinematics have provided a basis for interpreting descending commands from the motor cortex to muscles. More formally, we can define a resulting population response \(r(t)\) as a non-linear function \(f\) of external parameters \(m_1, m_2, ..., m_n\) as such
\begin{equation}
    r(t) = f(m_1(t), m_2(t),...,m_n(t))
    \label{eq:representational}
\end{equation}
Individual unit responses have been shown to encompass tuning to both types of movement parameters as well as time in a heterogeneous manner~\cite{churchland2007temporal, scott2008inconvenient, fetz1992recognizably}, making it increasingly difficult to reliably define the computations performed by the population - or even individual units - before or during movement. Recent progress in understanding the neural control of movement has instead been driven by a top-down approach: interpreting population-level mechanisms to further understand single unit responses, and ultimately, the overarching computation. This has led to a novel framework that interprets neural activity through the lens of a dynamical system; here, the activity or firing rates of individual neurons is hypothesized to form the \textit{states} of a high-dimensional dynamical system (\textbf{Figure \ref{fig2}} (A, B)). We denote this as the dynamical systems perspective \cite{2020vyasComputationNeuralPopulation}. Mathematically, we define a population trajectory \(h(t)\) and its derivative \(\dot{h}\), scaled by a time constant \(\tau\), with inputs \(s(t)\) and non-linear activation \(f\) using the following form.
\begin{equation}
    \tau \dot{h}(t) = f(h(t)) + s(t)
    \label{eq:dyn_persp}
\end{equation}
Equation \ref{eq:dyn_persp} describes the time varying activity of a population of recurrently connected non-linear units performing a computation leading to a kinematic trajectory through space. This framework clears uncertainties surrounding response complexity in individual units, given that some units may represent external parameters while others are responsible for shaping internal computations. However, as opposed to the model proposed in Equation \ref{eq:representational}, the nature of computations performed by a complex, high-dimensional dynamical system and their relation to movement generation are less obvious, thus requiring techniques to simplify and interpret the underlying signals. 

% Neural dynamics during reaching
\subsection{Manifolds during Motor Control}

How can we make sense of the heterogeneous, time-varying signals of a non-linear dynamical system and relate them to movement? It is possible that out of the many dimensions that make up the recorded neural space, only a small fraction are needed to sufficiently describe the system~\cite{gallego2017neural}. Given that the activity of neurons is likely constrained by the underlying circuitry connecting them, the effective dimensionality of the system is likely lower than that of the entire population. This gives rise to the notion that population activity may reside on a low dimensional surface embedded in a high dimensional population space, known as the manifold hypothesis (\textbf{Figure \ref{fig2}} (C)). 

In order to identify these manifolds, or subspaces, linear dimensionality reduction methods such as Principal Component Analysis (PCA), Factor Analysis, and Gaussian Process Factor Analysis have been key \cite{cunningham2014dimensionality}. These methods generally find the neural modes, or directions in neural state space, that define the latent variables of the population activity. The latent variables are the time dependent activity of the neural modes that can most sufficiently describe the signal. If only a few latent variables are needed to capture the most important features of the data (e.g., the variance), then we can view the neural modes as defining a hyperplane (or linear manifold) on which the activity resides. Indeed, such subspaces have been identified in motor and pre-motor cortices of primates performing center-out reaching tasks, with a three-dimensional manifold capturing target specific clusters of latent activity~\cite{santhanam2009factor}. The existence of such manifolds have been shown to impact the learning speed of primates controlling a cursor using a brain-computer interface (BCI), with task perturbations inside an intrinsic manifold being learned on a faster time-scale as compared to otherwise~\cite{sadtler2014neural}. The use of shared manifolds across distinct tasks with similar elements has also been discovered in primate M1~\cite{gallego2018cortical}. Low-dimensional manifolds are not limited to motor cortices during limb movements, and have additionally been discovered in a wide range of brain regions such as the pre-frontal cortex~\cite{mante2013context, kobak2016demixed}, V1~\cite{churchland2010stimulus}, olfactory cortex~\cite{kobak2016demixed}, and parietal cortex~\cite{raposo2014category} in various species such as monkeys and rats. 
 
 In traditional task structures involving a preparatory (or delay) epoch followed by movement, such as center-out reaching in primates and instructed directional licking in mice, activity predicting the upcoming movement has been shown to persist during the delay in the absence of external stimuli~\cite{tanji1976anticipatory, wise1985primate, churchland2006preparatory}. Activity during preparation and movement are shown to be nearly orthogonal~\cite{elsayed2016reorganization} (\textbf{Figure \ref{fig2}} (D)), thus occupying separate subspaces. This suggests the brain utilizes orthogonal manifolds to isolate distinct computations. Additionally, the preparatory subspace was found to be output-null~\cite{kaufman2014cortical}, now known as the null-space hypothesis~\cite{churchland2024preparatory}, explaining how such activity does not directly produce movement. This demonstrates the use of low-dimensional activity as a strategy employed by the brain to compartmentalize computations.

\subsection{Dynamics during Motor Control}

The time evolution (or dynamics) of neural activity, like the low-dimensional spaces they comprise, are constrained by network connectivity and shape neural computation, as recently demonstrated in BCI studies~\cite{oby2025dynamical}. Such dynamics have been studied during preparation, movement execution, and the transition between epochs in order to fully characterize the computations performed during delay-instructed tasks. To better observe preparatory dynamics, optogenetic stimulation of the mouse anterior lateral motor cortex (ALM) was performed during delay-instructed directional licking~\cite{inagaki2019discrete} (\textbf{Figure \ref{fig2}} (E)). Preparatory activity in this setting resembles a population ramp to threshold~\cite{inagaki2018low}. However, ramping may invoke various dynamical solutions, such as the state shifting along a continuous attractor, decaying to a discrete attractor, or moving in accordance with an externally driven discrete attractor. The results of the perturbations in~\cite{inagaki2019discrete} demonstrate that activity either rapidly recovers to its choice along a decision axis, or switches sides. The discrete nature of this shift, along with the rapid recovery of the state to pre-stimulation levels, suggests an externally-driven discrete attractor guiding the population state during preparation. Neural dynamics have also been studied in adjacent cognitive motor settings such as motor timing, which may utilize sequential activity or more complex population codes~\cite{paton2018neural, zhou2020neural}.

The next step in the preparation-to-execution pipeline involves the transfer of activity from one subspace to another. Large multi-phasic shifts in activity occur as preparation transfers to execution. In~\cite{churchland2012neural}, it was found that primates performing center out reaches displayed transient oscillatory dynamics after movement preparation. Large non-selective changes in activity have been shown to occur in response to the go-cue in primate M1/PMd, signaling movement onset itself~\cite{kaufman2016largest}. Such transient responses to the go-cue have also been shown in the mouse ALM~\cite{inagaki2022midbrain}, with the neural mode defining the go-cue response being the most prominent during memory-guided movement tasks. Such activity is thought to represent the shift of population activity from the null to output-potent movement subspace. 

During movement, it is believed that the motor cortex is tasked with generating coherent patterns of activity necessary to drive muscles. For example, primates performing cycling movements display elliptical neural trajectories in M1, likely generated by limit cycles~\cite{2022saxenaMotorCortexActivity} (\textbf{Figure \ref{fig2}} (F)). Such trajectories are stacked according to the speed at which the cycle is performed, with low trajectory tangling in comparison to muscle activity~\cite{russo2018motor}. The dynamical features of the subspaces that define movement preparation and execution have led to great strides in understanding how the brain prepares and executes movement.

\subsection{Role of Multiple Regions in Motor Control}

The motor cortex does not work in isolation to produce movement, but rather works in concert with other areas of the brain and body, including the basal ganglia, thalamus, cerebellum, and spinal cord. This results in a multi-regional circuit with distinct computational roles for each region likely depending on their underlying connectivity, cell-types, and functional specialization. Within the CBGTC loop, the role of the different pathways defined by striatal cell types has been extensively studied in relation to action selection~\cite{mink1996basal}. There are also direct reciprocal connections between the thalamus and cortex, studied in settings such as planning~\cite{guo2017maintenance, kao2021optimal} and sequencing~\cite{logiaco2021thalamic}, and between the thalamus and striatum, which may implement gating mechanisms~\cite{ding2010thalamic}. The STN also receives direct excitation from the cortex through the hyperdirect pathway, largely studied in relation to stopping signals~\cite{koketsu2021elimination}. The cerebellum has been shown to play a role in shaping the attractor landscape of the mouse ALM during preparation~\cite{li2020cortico}. While the cerebellum has primarily been studied in relation to internal models, evidence has shown that population level mechanisms such as a null-spaces are implemented as well~\cite{fakharian2025vector}. 

\subsection{Emulating Motor Control using Dynamical Systems }

Neural network models of non-linear dynamical systems, known as recurrent neural networks (RNNs), paired with gradient descent optimization of specified loss functions, have recently proved to be invaluable tools for testing hypothesis in both motor and cognitive settings. RNNs are a special class of artificial neural networks where each unit is recurrently connected with each other unit in the network. The units in RNNs directly model neuronal function: they integrate information from many inputs through weighted connections, and their outputs are governed by nonlinearities. This deep learning based modeling framework finds a set of weights, and consequentially a dynamical solution, to a specified objective function. When this objective is modeled after a laboratory experiment performed by a live animal, such as center-out reaching, the model provides a particular optimal solution to the task that can be compared with recorded neural data. This form of modeling is known as task-driven or goal-driven modeling \cite{yamins2016using, 2022saxenaMotorCortexActivity}. This is in contrast to data-driven modeling, where the RNN activity is directly constrained to the recorded neural data \cite{durstewitz2023reconstructing}. 

Formally, a commonly used form of RNNs is given by the equations below
\begin{equation} \label{eq:RNN}
\begin{split}
\tau \dot{x} &= f(h,s) = -x + W_hh+W_ss+b_h+\sqrt{2\tau \delta^2}\epsilon \\
 h & = \sigma(x) \\
 a &= W_o h + b_o
\end{split}
\end{equation}
Here, \(\tau\) represents the network time constant, \(W_h, b_h,\) and \(W_s\) represent the hidden weights, hidden bias, and input weights respectively, $\epsilon \sim \mathcal{N}(0,1)$, \(\delta\) is a noise scalar, \(\sigma\) is the non-linear network activation, and \(h\) and \(x\) represent the non-linear and linear RNN states respectively. RNNs in neuroscience typically include a linear readout \(a\), with \(W_o\) and \(b_o\) representing the output weights and bias respectively. Once trained in a goal-driven setting, researchers typically ``reverse-engineer'' the RNN to derive insight regarding how the task is solved \cite{sussillo2013opening}. Given the non-linear nature of the model, reverse-engineering RNNs typically involves linearizing about certain states in order to analyze the local dynamics. Typically, RNNs are linearized about fixed points, or the states \(h^*\) paired with input \(s^*\) such that \(f(h^*, s^*) = 0\), where \(f\) denotes equation \ref{eq:RNN}. The dynamics around fixed points are approximately linear and can be examined by performing a Taylor expansion of the network about the desired point. To do so, it is first necessary to identify fixed points by optimizing for a set of hidden states \(\{h^*_1, h^*_2, ...\}\) that minimize \(f(h^*, s^*)\), where \(s^*\) is a fixed input~\cite{sussillo2013opening}. Once fixed points are captured, the network can be linearized about the fixed point \(h^*\) as such
\[f(h^*+\delta h^*, s^* + \delta s^*) \approx f(h^*, s^*) + \frac{\partial f}{\partial h}(h^* s^*)\delta h + \frac{\partial f}{\partial s}(h^*, s^*) \delta s\]
By definition, \(f(h^*, s^*) = 0\), and second order terms are approximately zero given that \(||\delta h||^2 \approx 0\). Additionally, assuming the input \(s^*\) is held constant, we can ignore any changes from \(\delta s\). Thus, our desired system simplifies to
\[f(h^*+\delta h^*, s^* + \delta s^*) \approx \frac{\partial f}{\partial h}(h^*, s^*)\delta h\]
The eigenvalues, phase portraits, and other tools common to analysis of linear systems can then be applied to the resulting Jacobian to interpret network computation. 

The above methods have been predominantly used for cognitive tasks given their dependence on fixed point computations. The first use of reverse engineering in RNNs to guide experimental analysis was shown in~\cite{mante2013context}, where RNNs trained to mimic the prefrontal cortex performing a context based color-motion discrimination task utilized a continuous attractor to integrate evidence while changing the direction of its velocity field based on context. In~\cite{stroud2023optimal}, reverse engineering of network dynamics was utilized to determine optimal input directions for performing a working memory task. Network dynamics have additionally been explored in multitasking frameworks, where shifts in dynamics across tasks were observed~\cite{yang2019task, driscoll2024flexible}. Dynamical solutions found by networks have been shown to be consistent across architectures as well~\cite{maheswaranathan2019universality}.

For motor control, the focus has been primarily on modeling the motor cortex generating patterns of activity necessary to drive muscles. RNNs that reproduce EMG data have been shown to closely match recorded M1 activity when the network is incentivized to find simple solutions using specialized regularizations~\cite{sussillo2015neural}. Networks trained to produce muscle activity for monkeys performing cycling movements at different speeds have been shown to incorporate elliptical trajectories stacked along a particular network mode, in line with experimental findings~\cite{2022saxenaMotorCortexActivity}. In~\cite{feulner2025neural}, it was demonstrated that an RNN can utilize an error-based feedback signal to adapt to perturbations such as visuomotor rotation. RNNs with biologically inspired constraints such as Dale's law and excitatory-inhibitory balance have been used to test hypotheses regarding the underlying dynamics controlling primate reaches~\cite{o2022direct}. These tools continue to guide our theories of cortical control of movement in an experimentally verifiable manner.

% Heading 1: OPTIMAL CONTROL AND DEEP REINFORCEMENT LEARNING FOR UNDERSTANDING THE NEURAL CONTROL OF MOVEMENT
\section{OPTIMAL CONTROL THEORY FOR UNDERSTANDING THE CONTROL OF MOVEMENTS}
\label{ch:opt_ctrl}
\begin{figure}[h]
\includegraphics[width=0.8\textwidth]{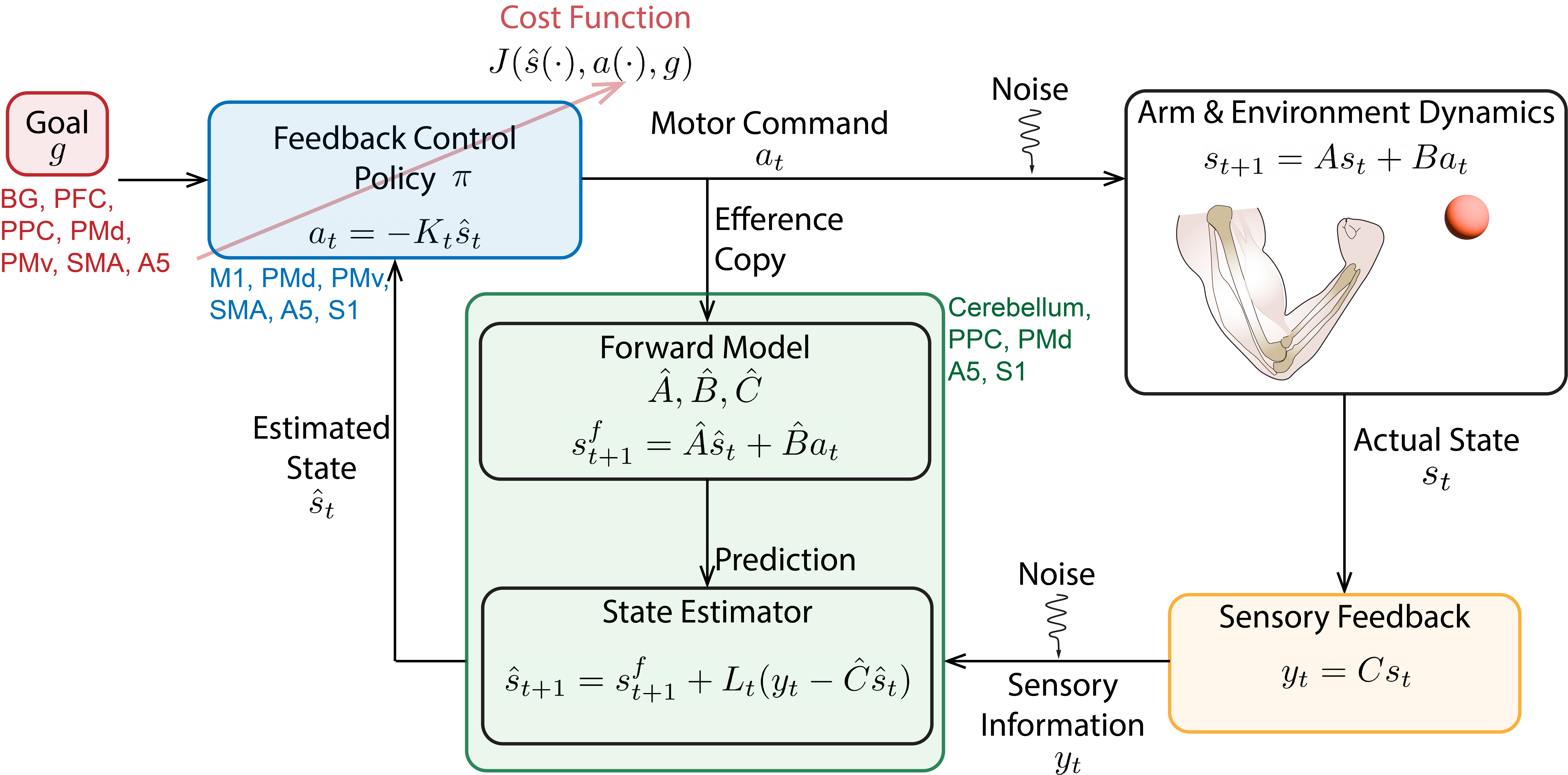}
\caption{{\bf The framework of optimal control applied to sensorimotor control.} The arm and environment dynamics are approximated to be linear with state $s$, and a forward model and state estimator are hypothesized to compute an estimated $s$. The feedback controller receives the estimated state and a goal in order to compute a feedback control policy $\pi$ given a cost function $J$. The hypothesized brain regions where these specific computations take place are included in different colors (BG: basal ganglia; PFC: prefrontal cortex; PPC: posterior parietal cortex; PMd: dorsal premotor cortex; PMv: ventral premotor cortex; SMA: supplemental motor area; A5: area 5; M1: primary motor cortex; S1: primary somatosensory cortex). Adapted from \cite{takeiTransientDeactivationDorsal2021,scott_computational_2012}.}
\label{fig:opt_ctrl}
\end{figure}

While dynamical models of recorded neural activity have been extensively explored (see Section \ref{ch:neur_dyn}), there has been relatively little effort to understand the computational goal of these dynamics, such as the optimal control of limb dynamics. In fact, sensorimotor control has vastly benefited from being cast in an optimal control framework \cite{todorov2002optimal,scott2004optimal, liu2007evidence, franklin2011computational, saxena2020performance}. The theoretical framework of optimal control formalizes the concept of the brain’s reliance on sensory feedback while achieving a desired goal (\textbf{Figure \ref{fig:opt_ctrl}}). Here, usual formulations posit that behavioral dynamics operate according to linear gaussian models. While the control of behavior is thought to be implemented by different regions of the brain, this assumption is not explicitly reflected in the formulation. The optimal control solution comprises of state estimation and a control policy, which can be determined using a user-defined cost function. 

\subsection{Internal Models} 
The transformation from motor commands to sensory feedback is governed by the dynamics of the musculoskeletal system and the physical world. Although such processes are executed in the external physical world, the brain is hypothesized to construct an internal model to represent this transformation \cite{jordan1996handbook, kawato1987hierarchical}. The internal processes of the brain that model this aspect of the transformation are known as internal models, or forward models. These models are thought to predict the next state of the environment given the current state and a copy of the motor command. Evidence suggests that such internal models are primarily implemented in the cerebellum \cite{wolpert1998internal}. An integrative theoretic account \cite{miall1996forward, therrien2015cerebellar} suggests that lack of motor coordination and stability can result from absence of internal predictive feedback and that cerebellum contains internal models that are crucial to overcome such behavioral deficits \cite{kawato2003internal, diedrichsen201438, muller1994dyscoordination, miall2007disruption}. 

In the probabilistic framework, the forward models $p_{f}$ encode the probability distribution over the possible future states $s_{t+1}$ given the current state $s_t$ and the motor command $a_t$:

\begin{align}
    p_f(s_{t+1} \vert s_t, a_t)
\end{align}
 
A prediction of the future state trajectory, $s= \{s_1, ...., s_T\}$, given the action trajectory, $a= \{ a_0, ..., a_{T-1} \}$, can be obtained by the repetitive application of the forward models:

\begin{align}
    p(s \vert s_0, a) = \prod_{k= 1}^{T} p_f( s_k \vert s_{k-1}, a_{k-1} )
\end{align}

This prediction of future state trajectory is particularly useful in motor planning. It is hypothesized that the brain also encodes priors over the sensory signals $p(y)$ and the motor signals $p(a)$, reflecting its belief about these variables before any actual sensory feedback is received. Such internal models are also known as prior models and several studies point towards their existence \cite{fiser2010statistically}. Prior models in combination with the forward models can be used to formulate inverse models. 

The internal processes that compute the optimal motor commands given a desired environmental state are known as inverse models. As the output of the inverse models are muscle excitations that produce the desired consequences in the external environment by controlling the musculoskeletal model, we use the terms ‘inverse models’ and ‘controller’ interchangeably. Evidence suggests that the implementation of the inverse models may be distributed among several brain regions, such as in cerebellum \cite{wolpert1998internal} and in motor cortex \cite{wolpert1998internal, arce2010combined}. 

Consider, for example, the problem of computing an optimal action trajectory $a^*$ that generates a movement towards a goal state $g$. Combining forward models with Bayesian inference can be used to determine the joint probability distribution of the state and action trajectory given the observation of goal state $g$: 

\begin{align}
    p( s, a \vert g ) = p_f(s \vert s_0, a) p(g \vert s) p (a)
    \label{neq1}
\end{align}

An inverse model is then a mapping from the desired goal state to the action that can be obtained from \eqref{neq1} by integrating out $s$:

\begin{align}
    p_{\mathrm{inv}}(a \vert g) = \int_{s} p_f (s \vert s_0, a) p(g \vert s) p(a) ds
\end{align}

The optimal action $a^*$ maximizes the distribution specified by the inverse model:
\begin{align} 
    a^* = \mathrm{arg}\max_{a} \ p_{\mathrm{inv}} (a \vert g) 
\end{align} 

Much of the complexity of associated neural processes and ensuing behavior arises from the interactions between inverse and forward models. Next, we will show how internal models play a crucial role in all aspects of sensorimotor integration and control: a complex process through which the brain uses sensory feedback from the external environment and internally generated task-relevant signals for motor learning, planning, and control. 

\subsection{Motor Learning}
Properties of the sensorimotor system change at different timescales, for example, on a short timescale, involving interactive processes with the external environment, and on a longer timescale, due to evolutionary processes such as growth. Internal models must adapt continuously to account for these changes. The learning of forward models is relatively straightforward using the error between the predicted and the actual sensory feedback. The neural mechanisms underlying such predictive learning have been studied in several model systems, such as the cerebellum-like structure of the electric fish \cite{bell1997synaptic}. 

Acquiring inverse models is generally more involved, primarily due to the sensory-to-motor coordinate transformation required in the computation of appropriate gradients. When a movement is made, the sensorimotor system can sense the directional error between the resulting and the desired sensory outcome. However, the sensorimotor system needs to convert this sensory prediction error from sensory coordinates into appropriate gradients required to update each element of the motor command. Evidence suggests that the sensorimotor system is highly efficient in learning the gradient of the sensory prediction error with respect to the changes in motor commands even when the mapping from sensory to motor coordinates is perturbed \cite{mosier2005remapping, johansson2006lateralized, liu2011reorganization}. There is extensive evidence that error-based learning characterized by fast adaptation depends on the cerebellum \cite{tseng2007sensory, golla2008reduced}. In addition to error-based learning, reward-based reinforcement learning (RL) is particularly useful when a sequence of actions is needed to solve a motor task and the outcome is far removed from a particular action \cite{izawa2011learning, abe2011reward}.

\subsection{State Estimation}

To construct inverse models, the sensorimotor system needs information about the current state of the environment. However, it faces three main challenges. First, biological sensorimotor loops are slow and involve significant sensory delays. Second, noise contaminates various stages of the sensorimotor loop, i.e., motor outputs and sensory inputs from the environment. Third, the sensory inputs from the environment may provide only partial information about its state. All these factors make online control impractical while carrying out most complex and fast movements. To overcome these challenges, the sensorimotor control system is hypothesized to use a combination of forward models and actual sensory feedback from the environment to estimate its state in an observer framework. In the case of linear systems, the Kalman filter is an optimal observer as it estimates the state with the least squared error \cite{stengel1994optimal}. The Kalman filter model is a combination of two processes. In the first process, this model uses the efference copy of the motor command and the current state estimate to generate the next state estimate using the internal forward model. In the second process, the difference between actual and expected sensory feedback is used to refine the next state estimate. The relative weighting between these two processes is modulated optimally by the Kalman gain, $L_t$. 

\begin{align}
    \hat{s}_{t+1} = \hat{A} \hat{s}_t + \hat{B} a_t + L_t (y_t - \hat{C} \hat{s}_t)
\end{align}

where $\hat{A}$, $\hat{B}$, and $\hat C$ consist of the forward model: they are the brain's estimates of the arm and environment's dynamics (\textbf{Figure \ref{fig:opt_ctrl}}). When the environmental dynamics are non-linear or the sensory noise is non-Gaussian, linear approximation approaches such as Extended Kalman Filters, unscented filters, or particle filter can also be used \cite{stengel1994optimal, todorov2006optimal}. 

The observer framework serves a variety of roles in biological motor control, such as sensory reafference cancellation, forward state estimation or mental simulation of intended movements, prediction for learning and planning novel behaviors, to name a few. Several empirical studies have investigated the existence of such estimates \cite{van1999integration, merfeld1999humans,mulliken2008forward}.

\subsection{Motor Planning and Control} 
Motor tasks are usually specified at a high-level, such as reaching for a cup of coffee. However, the sensorimotor system must work at a detailed level, specifying the activations for each of the relevant muscle, that are in turn converted into the excitations, joint torques and finally to the path of the hand in space. A given motor task can be achieved in infinitely different ways. Consider, for example, all the possible hand paths with which to reach for the cup of coffee. Given all the redundant ways to achieve a motor task, it is surprising that the sensorimotor system generates remarkably stereotypical behaviors: both within the repetition of the same task and between individuals on the same task. Optimal control provides an elegant framework to deal with such selection problems. Cost functions provide a criterion with which to evaluate all the different possible movements, including successful movement execution to the goal state $g$, as well as enforcing constraints such as minimizing muscle effort. Cost functions are usually specified as functions of the state (movement), motor command (actions), and the goal or task. 

Optimal control models have been proposed based on maximizing the smoothness of the joint torques (minimum torque-change) \cite{uno1989formation} and hand trajectory (minimum jerk) \cite{flash1985coordination} for arm movements. Optimal control models based on signal-dependent noise have provided a unifying cost function for goal-directed eye and arm movements \cite{harris1998signal}. 
Todorov and Jordan \cite{todorov2002optimal} deployed stochastic optimal control with energy-minimization constraints to show that the nervous system may correct movements in task-relevant dimensions while allowing for high variability in task-irrelevant dimensions, known as the minimum intervention principle. 
However, one of the challenges of the field has been to design a unified cost function that can explain a large repertoire of movements in dynamic settings, while being based on quantities that are plausibly important to the nervous system and can be directly or indirectly measured. 

\subsection{Algorithms for Optimal Control}
\label{ch:ofc:algs}
Here, we provide two algorithmic solutions for optimal control that are commonly used in the frameworks described above.
\subsubsection{Dynamic Programming}
Cost functions $c(s_t, a_t)$ as a function of the state $s \in \mathcal {S}$ and action $a \in \mathcal{A}$ are usually specified at each timestep, $t$. %Here, $s \in \mathcal {S}$ is the state and action $a \in \mathcal{A}$ is the motor command. 
The goal of the controller is to minimize the cumulative cost, $J(s(\cdot), a(\cdot)) = \sum_{t = 0}^{T} c(s_t, a_t)$ incurred over the entire movement trajectory. However, it is not possible to compute the current optimal action $a^{*}_t$ without knowing its future consequences. Dynamic Programming (DP) is used to solve such sequence-based optimal control problems. DP is based on Bellman's optimality principle, which states that any part of the optimal state-action sequence is also optimal. This allows for solving the optimal control law or policy, $\pi: \mathcal{S} \rightarrow \mathcal{A}$, recursively by starting from the final state and working backwards to the initial state. For notational clarity, the estimated state $\hat{s}$ evolution dynamics in this section are given by: $s_{t+1} = f(s_t, a_t)$. 

The key to DP is the optimal value function which captures the long term consequences of an action, by calculating the minimum cost-to-go for a given state. The optimal value function is defined as: 

\begin{align}
    v(s) = \mathrm{min}_{a \in A(s)} \{ c(s, a) + v (f(s, a)) \}
    \label{neq2}
\end{align}

For a given state $s$, the value function $v$ represents the minimum cost that will be incurred to reach the target state $s_T$ starting from $s$. Although the optimal value function captures long-term consequences of an action, \eqref{neq2} enables its computation in a greedy manner (using only the local information): we need to consider only the immediate cost of every possible action and add to it the optimal value of the resulting/next state. The optimal control law $\pi$ is computed as follows: 
\begin{align}
    \pi(s)= \mathrm{argmin}_{a \in A(s)} \{c(s, a) + v(f(s, a))  \}
    \label{neq3}
\end{align}
Equations \eqref{neq2} and \eqref{neq3} are also known as Bellman equations. 

If we know the optimal values of all the resulting states possible from a given state $s$, we can use \eqref{neq3} to compute the optimal control law $\pi$. DP thus provides a useful approach to compute $\pi(s)$ and $v(s)$. The key is to start from the target or absorbing states for which the optimal values or the final costs are given. Then, using equations \eqref{neq2} and \eqref{neq3}, perform a backward pass in which every state is visited after all its successor states have been visited. Value iteration and policy iteration are similarly based on iteratively improving the initial guesses of the value functions and are guaranteed to converge to optimal solution. RL is another method to solve such discretized optimal control problems and relies on exploration for state visitation, for example, using a stochastic policy. 

\subsubsection{Linear Quadratic Gaussian}
Now, we turn to continuous time stochastic optimal control problems that yield closed form solutions for the optimal feedback control policy. Consider the following environmental dynamics: 

\begin{align}
    ds = (As + Ba)dt + Fdw
\end{align}

where, $w$ represents the Brownian motion. 

Let the associated quadratic instantaneous $c$ and final costs $h$ be: 

\begin{align}
    c(s, a) = \frac{1}{2}s^{T}Qs + \frac{1}{2} a^{T}Ra
\end{align}

\begin{align}
    h(s) = \frac{1}{2}s^{T}Q^{f}s
\end{align}

where $Q$ and $Q^f$ are symmetric matrices, and R is a symmetric positive-definite matrix. 

The optimal action $a^*$ is given by the following control policy $\pi$:

\begin{align}
    a^* = -Ks = - R^{-1}B^{T}V_ts
    \label{neq4}
\end{align}

with $\frac{d}{dt}V_t = \dot{V}_t$ given as:

\begin{align}
    -\dot{V_t} = Q + A^{T}V_t + V_tA - V_tBR^{-1}B^{T}V_t
    \label{neq5}
\end{align}

With the boundary conditions $V(T)= Q^f$. The ODE \eqref{neq5} can thus be solved backward in time to compute the function $V$. %The optimal control law $\pi$ is given by \eqref{neq4}. 
In the case of deterministic environmental dynamics ($F=0$), the optimal policy remains the same - known as the linear quadratic regulator (LQR). 

Optimal control algorithms are typically limited to generating simple movements and, importantly, lack an explicit neural representation of the feedback control policy, relying instead on optimization methods. Even in nonlinear settings such as in \cite{todorov2005generalized}, locally optimal actions are typically computed using time-varying linear functions of the estimated state. Consequently, this framework has generated limited neural predictions with some exceptions \cite{scott_computational_2012, ueyama2017optimal}. 
Moreover, biological evidence suggests that exploration plays a crucial role in motor learning and generalization \cite{sokoloff2020spatiotemporal}. However, optimal control algorithms compute optimal actions analytically and do not rely on exploration. The next section instead turns to RL for neural predictions using embodied control. 

%%%%%%%%%%%%%%%%%%%%%%%%%%%%%%%%%%%%%%%%%%%%%%%%%%%%%%%%%%%%%
% Combining neural population dynamics with embodied control%
%%%%%%%%%%%%%%%%%%%%%%%%%%%%%%%%%%%%%%%%%%%%%%%%%%%%%%%%%%%%%
\section{SIMULATING EMBODIED CONTROL FOR ELUCIDATING NEURAL CONTROL OF MOVEMENT}
%[Improve introduction to better tie together previous to sections]

\begin{figure}[h]
\includegraphics[width=0.95\textwidth]{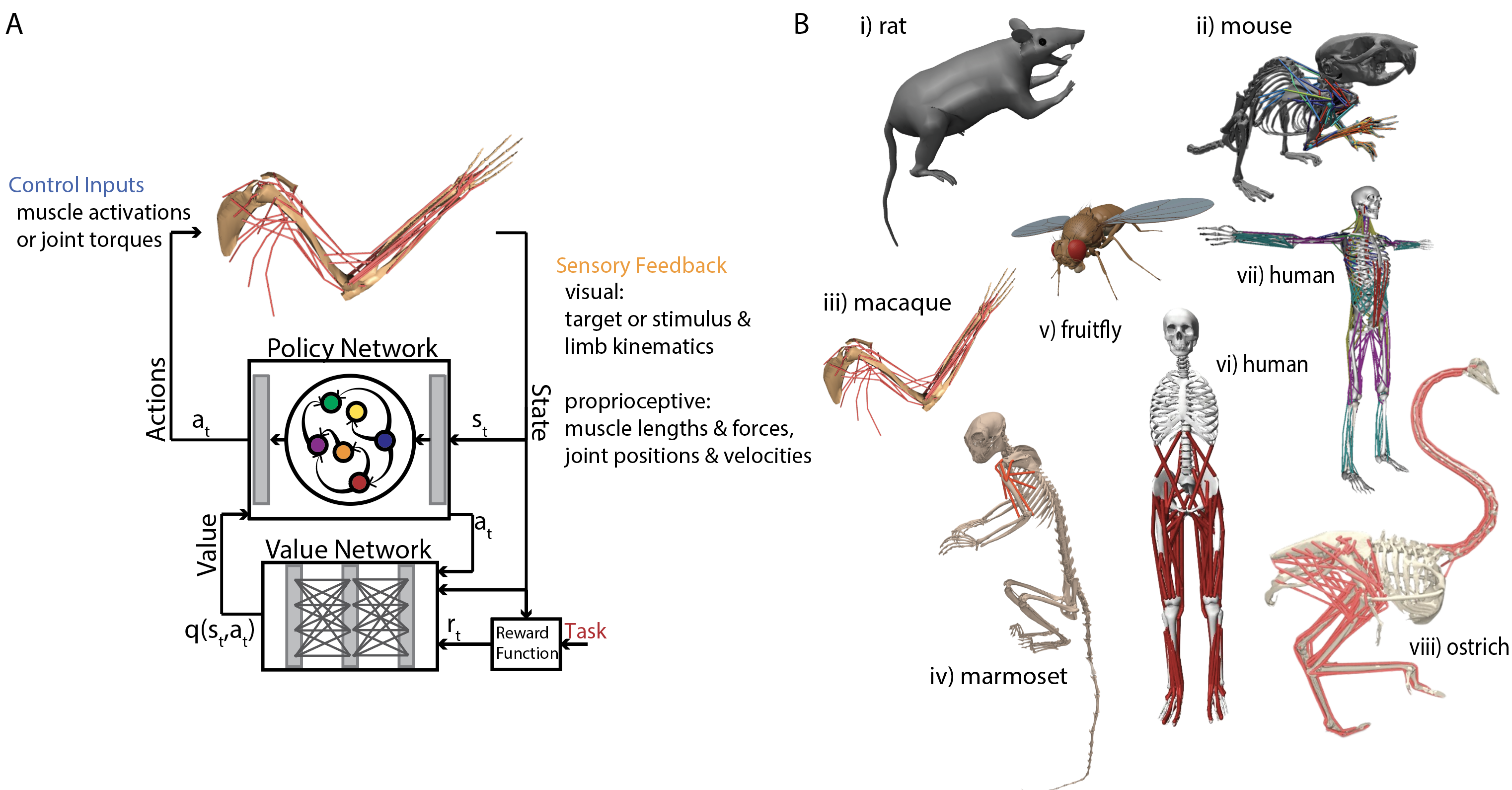}
\caption {{\bf Deep reinforcement learning with simulated bodies for embodied control.} A) Actor critic deep reinforcement learning setup for modeling musculoskeletal control. Adapted from \cite{2024almaniMSimGoaldrivenFramework}. B) Menagerie of physically simulated models: i) rat \cite{2024aldarondoVirtualRodentPredicts, merelDeepNeuroethologyVirtual2019b}, ii) mouse \cite{dewolf_neuro-musculoskeletal_2024}, iii) macaque arm \cite{chan2006computational, 
2024almaniMSimGoaldrivenFramework}, iv) marmoset \cite{walker_building_2023}, v) fly \cite{2025vaxenburgWholebodyPhysicsSimulation}, vi) human lower body \cite{2021songDeepReinforcementLearning}, vii) human full body \cite{nakada_deep_2018}, viii) ostrich \cite{labarberaOstrichRLMusculoskeletalOstrich2022}. Adapted from \cite{merel_hierarchical_2019, dewolf_neuro-musculoskeletal_2024,2024almaniMSimGoaldrivenFramework, walker_building_2023,  2021songDeepReinforcementLearning, nakada_deep_2018, labarberaOstrichRLMusculoskeletalOstrich2022}}.
\label{fig4}
\end{figure}

Dynamical models of the brain in the form of RNNs, as well as optimal control formulations, have both enhanced our knowledge of motor control, albeit from parallel perspectives. In reality, both processes are necessary in order to fully characterize the computations underlying the neural control of movement, however the synergy between both frameworks is relatively unexplored. Progress has recently been made on this front through the use of RNNs in feedback with skeletal or musculoskeletal models performing a variety of behaviors, trained using deep RL (DRL) (\textbf{Figure \ref{fig4}A}). This framework, which we term embodied control, has the potential to bridge the gap between the dynamical systems perspective of neural computations and the role that feedback plays from a control theoretic perspective.

\subsection{Musculoskeletal Models}
Bodies - specifically, musculoskeletal systems - are critical components of the dynamical machinery that generates complex vertebrate behavior.  Any holistic description of motor control must therefore include an explicit description of musculoskeletal systems and their interaction with neural circuits. There is a rich history of modeling musculoskeletal control in the field of biomechanics, which has produced a variety of biomechanical models and physics simulators for control simulation (e.g., OpenSim, see \cite{seth_opensim_2011}).  There have also been efforts toward framework compatibility through tools like myoconverter \cite{ikkala2022converting}, which allows one to convert an Opensim model into formats compatible with other engines (e.g., Mujoco \cite{todorov_mujoco_2012}). Mujoco is a general purpose physics engine that achieves fast simulation of muscle dynamics due to its simplified muscle model, and has proved useful in this regard \cite{caggiano_myosuite_2022, wang2022myosim}. Related ecosystems now include packages for neuromusculoskeletal optimization (e.g., Moco in OpenSim \cite{dembia_opensim_2020}) and emerging differentiable or real-time simulators (e.g., Brax \cite{freeman_brax_2021}) that aim to close the loop between control theory and embodied implementation.  This expanding ecosystem of tools and approaches provides the context for current efforts to both understand and engineer embodied control of complex behavior.  

Incorporating musculoskeletal models in a comprehensive model for motor control has inherent challenges associated with it, since there is significant complexity on multiple spatial and temporal scales in musculoskeletal systems.  For example, pennation angles and muscle architecture shape the 3D structure of force production, in cases such as the pectoralis major, and this makes modeling muscles with unidirectional contractile units a significant simplification. In addition,  heterogeneity in muscle fiber type is clearly functionally significant in biology \cite{schiaffino_fiber_2011}, resulting in muscles with variable activation dynamics and resistance to fatigue, often completely ignored in musculoskeletal modeling efforts. Inaccurate body models can distort estimates of neural control signals and bias conclusions about the principles of coordination. 

\subsection{Deep Reinforcement Learning to simulate Embodied Control}
During the last decade, researchers have begun to look to DRL to model locomotion and dexterous manipulation using musculoskeletal models \cite{2021songDeepReinforcementLearning, wang2022myosim}. 
DRL is a subfield of machine learning that deals with discretized optimal control problems and may address the shortcomings of optimal control theory as mentioned at the end of Section \ref{ch:opt_ctrl}. Here, the sensorimotor loop is modeled as a controller, parameterized using a neural network $\theta^{\pi}$, interacting with the environment (\textbf{Figure \ref{fig4}A}).

Many fundamental concepts in RL have their analog in optimal control theory. Instead of cost functions, RL is based on reward functions, $r(s_t, a_t)$. In RL, the long-term consequences of an action are usually captured by the action-value function $Q$ (much like its counterpart value function in optimal control): 
\begin{align}
    Q^{\pi}(s_t, a_t) = \mathop{\mathbb{E}}_{r_t, s_{t+1} \sim E} [r(s_t, a_t) + \gamma \mathop{\mathbb{E}}_{a_{t+1} \sim \pi}[Q^{\pi}(s_{t+1}, a_{t+1}) ] ]
\end{align}
where, $\gamma \in [0, 1]$ is known as the discounting factor, and the rest of the notation follows from Section \ref{ch:ofc:algs}. 

While modeling sensorimotor control, an additional neural network parameterized by $\theta^{Q}$ is typically included for learning the action-value function. Dopaminergic projections to the motor cortex can constitute a possible neural correlate of reward functions \cite{luft2009dopaminergic}.  
The goal is to find a feedback control policy that maximizes the cumulative return (analogous to minimizing the cumulative cost in optimal control): 

\begin{eqnarray}
    \nabla_{\theta^{\pi}} J &=& \mathop{\mathbb{E}}_{s_{t} \sim \rho^{\beta}} [  \nabla_{\theta^{\pi}} Q (s, a \vert \theta^{Q} )  \vert _ { s = s_t, a= \pi(s_t \vert \theta^{\pi}) } ] \\
    &=& \mathop{\mathbb{E}}_{s_{t} \sim \rho^{\beta}} [  \nabla_{a} Q (s, a \vert \theta^{Q} )_ {  \vert s = s_t, a= \pi(s_t )  } \nabla_{\theta_{\pi } } \pi(s \vert \theta^{\pi}) \vert_ {s= s_t} ] 
\end{eqnarray}

This is also known as policy gradient as discussed in Silver et al. \cite{silver2014deterministic}. $\rho^{\beta}$ reflects the state visitation distribution under a different policy $\beta$ and is used to emphasize off-policy learning, such as learning from past experiences in biological motor control. 

\subsection{Towards embodied control of complex behavior}
Researchers motivated to engineer general motor intelligence for whole-body humanoid control have been developing training strategies and architectures towards such flexibility largely within the general purpose MuJoCo physics engine \cite{todorov_mujoco_2012,merel_catch_2020}.  
Although complex behavior can emerge through exploration alone \cite{heess_learning_2016}, these learned solutions may look unnatural.  However, in the last few years, a number of studies have applied imitation learning from motion capture data to agents of various embodiments to model diverse behaviors \cite{merel_learning_2017, peng_deepmimic_2018}. Briefly, imitation learning consists of a family of RL methods where an agent learns a policy by copying expert behavior instead of explicitly optimizing a reward function through trial and error \cite{bohez_imitate_2022}. Enabled by recent progress in deep-learning-based marker-less motion capture, which has enabled rich quantification of animal movement for neuroscience applications \cite{nath_using_2019,pereira_sleap_2022}, this strategy has proven useful to neuroscience through the modeling of naturalistic behaviors captured during experiments in a diverse range of species (e.g. rats \cite{2024aldarondoVirtualRodentPredicts}, mice \cite{dewolf_neuro-musculoskeletal_2024}, and flies \cite{2025vaxenburgWholebodyPhysicsSimulation} (\textbf{Figure \ref{fig4}B})).

Some of the most impressive examples of this work come from multiple groups independently developing whole-body fruitfly models capable of recapitulating fly behaviors such as locomotion, flight, and odor plume tracking \cite{lobato-rios_neuromechfly_2022, wang-chen_neuromechfly_2024,2025vaxenburgWholebodyPhysicsSimulation} (\textbf{Figure \ref{fig4}B(v)}). The controllers used across these projects vary, though all incorporate central pattern generators modulated by top-down hierarchical architectures. These hierarchical networks are biologically inspired analogs of the fruitfly motor system, where the fly brain operates on the body through the ventral nerve cord. \cite{karashchuk_sensorimotor_2025} use this framework to test the effect of sensorimotor delays in locomotor stability in a whole body fly model.

Not surprisingly, there are a growing number of efforts to build DRL-driven control of human musculoskeletal models, bolstered by community challenges involving locomotion and manual dexterity \cite{2021songDeepReinforcementLearning} (\textbf{Figure \ref{fig4}B(vi)}). One such challenge involved rotating Baoding balls within the palm of a human musculoskeletal hand model. The winning solution to this challenge \cite{chiappa_acquiring_2024} used curriculum learning and DRL to obtain impressive dexterous control, and found that the learned solutions were consistent with the muscle synergies used by human subjects. In another example, \cite{simos_reinforcement_2025} develop an imitation learning framework capable of controlling a whole body skeletal model with lower limb musculature. The corresponding muscle activity patterns of the model correlate well with those of humans engaged in locomotion. It is worth noting that impressive whole body musculoskeletal control can also be obtained through supervised learning \cite{nakada_deep_2018} (\textbf{Figure \ref{fig4}B(vii)}), and DRL-driven imitation has been obtained in non-traditional model species (e.g. \cite{labarberaOstrichRLMusculoskeletalOstrich2022} (\textbf{Figure \ref{fig4}B(viii)})).

\subsection{Neural representations and dynamics in embodied systems}

Recent work has furthered the utility of embodied control models to compare the representations of RNNs driving behavior with neural data. Merel et al. \cite{merelDeepNeuroethologyVirtual2019b} developed a complete rat body model (\textbf{Figure \ref{fig4}B(i)}) and trained it using DRL to perform multiple tasks. They then used approaches derived from neuroscience to characterize the learned representations within the value and policy networks, such as revealing rotational dynamics during behavior~\cite{churchland2012neural}. In subsequent work,  \cite{2024aldarondoVirtualRodentPredicts} used whole-body kinematics recorded from rats engaged in open field behaviors and trained a virtual rat on the same behaviors using imitation learning.  Through comparison of neural recordings from rat motor cortical and dorsal striatum populations with the policy and value network activity, the authors found that the activity of the inverse dynamics model realized by the policy and value networks was a better fit to experimentally recorded activity than that of alternative representational models. In addition to a rat, multiple groups have developed whole body mouse models \cite{tata_ramalingasetty_whole-body_2021, dewolf_neuro-musculoskeletal_2024} (\textbf{Figure \ref{fig4}B(ii)}). In recent work \cite{dewolf_neuro-musculoskeletal_2024} used a whole body mouse skeletal model with upper limb musculature to investigate the coordinate sytems encoding sensorimotor prediction errors in recorded populations from M1 and S1  during motor  adaptation.

Despite the relatively early development of a whole body skeletal model of a macaque monkey \cite{putrino_training_2015}, much of the recent work to develop DRL-driven musculoskeletal modeling of macaques has focused on the upper limb \cite{2022almaniRecurrentNeuralNetworks, 2024almaniMSimGoaldrivenFramework} (\textbf{Figure \ref{fig4}B(iii)}). In \cite{2024almaniMSimGoaldrivenFramework, 2022almaniRecurrentNeuralNetworks}, Almani et al. developed a framework called MuSim, which uses an actor-critic framework (\textbf{Figure \ref{fig4}A}) to train RNNs in feedback with a macaque upper limb model to reproduce cycling behaviors performed by primates during experiments. In addition to reliably recapitulating the target kinematics, and demonstrating strong correlations between RNN and recorded neural activity on trained behaviors, they show that neural activity in unseen conditions could be predicted by the model. Such generalization was likely aided by the use of explicit neural constraints while training the policy, similar to~\cite{sussillo2015neural} (\textbf {Box \ref{sidebar:regularization}}). The inherent reliance on exploration in DRL may also contribute to the model's generalization capabilities. In subsequent work, the authors present a broader framework for modeling both musculoskeletal dynamics and recorded neural activity by incorporating a semi data-driven approach in addition to network constraints \cite{2024almaniMSimGoaldrivenFramework}. In addition to macaques, common marmosets are growing in prominence as a primate model well-suited for studying complex natural behaviors requiring feedback and prediction \cite{shaw_fast_2023, walker_building_2023, walker_chronic_2021} (\textbf{Figure \ref{fig4}B(iv)}). Future work involves training such embodied models to perform a diversity of complex natural behaviors.

In fruitflies, efforts to align networks with specific details of fruitfly neuroanatomy or neural recordings are in early stages. One example is \cite{wang-chen_neuromechfly_2024}, who used a connectome-constrained approach to model the fly visual system\cite{lappalainen_connectome-constrained_2024} performing object detection as part of an embodied simulation of courtship behavior. Advances in drosophila connectomics \cite{dorkenwald_flywire_2022} offer tremendous potential to build accurate sensorimotor circuit models, combining whole body models with connectome-derived network architectures. Such comparisons between experimentally recorded and simulated neural circuits in fruitfly remains a promising opportunity for future work.

\begin{textbox}[h]\section{NEURAL CONSTRAINTS IN EMBODIED CONTROL} 
Neural constraints, such as limitations on firing rates due to energetic costs or refractory periods, undoubtedly shape the solution space available to organisms for sensorimotor control. However, such constraints cannot be enforced in optimal control models as in Section \ref{ch:opt_ctrl}, as they lack an explicit neural implementation of policy. In the context of DRL-driven embodied control, biophysically-relevant neural and energetic constraints can be implemented using regularizations on the policy (and value) networks and reward functions. Such constraints enable DRL-driven models of embodied motor control to explain a broad repertoire of behavioral and neural data. Constrained models can generalize to produce movement trajectories unseen during training, as well as explain the corresponding neural data \cite{2024almaniMSimGoaldrivenFramework}.
Specifically, \cite{2024almaniMSimGoaldrivenFramework} considered the following regularizations for an RNN-based policy network:

\subsection{1 - Encourage low firing rates}
\begin{center}
\(  R_{FR} = \sum_{c=1}^{C} \sum_{i=1}^{N} \int_{0}^{T} h_{i} (c, t) ^ {2} dt \) 
\end{center}
where $C$ is the number of training conditions with $T$ timesteps per condition. $h$ is the activity or `firing rates' of the RNN units, in a network with $N$ total units.

\subsection{2 -  $\mathcal{L}_2$ penalty on input and output weights}

\begin{center}
    \(  R_{L2} = \|W_s\|_F^2 + \|W_o\|_F^2   \)
\end{center}

where \(W_s\) is the the input weight matrix and \(W_o\) is the output weight matrix as in Equation \ref{eq:RNN}, and \(\|\cdot\|_F\) denotes the Frobenius norm. This is a commonly used loss term for regularizing network weights.

\subsection{3 - Encourage simple population dynamics}

\begin{center}
   \( R_{W_h} = \frac{1}{CT}\sum_{c=1}^{C} \int_{0}^{T} \| \frac{\partial (W_hh(c, t))}{\partial x(c, t)} \|_F^2 dt \)
\end{center}

where $W_h$ represents recurrent weights and $x$ represents the activity of RNN units before the non-linearity. This loss term was originally used in \cite{sussillo2015neural} and has been found to be very helpful for the emergence of neural-like solutions.

\subsection{4 - Data-Driven Modeling}
Here, a subset of policy network units are constrained to the recorded neural data:

\begin{center}
    \( R_{D} = \sum_{i=1}^{N_{REC}} \sum_{c=1}^{C} \int_{0}^{T} (h_i(c, t) - n_i (c, t)) ^ 2  dt \) 
\end{center}

where $n$ represents the firing rates of recorded neurons and $N_{REC}$ is the number of recorded neurons.

\label{sidebar:regularization}
\end{textbox}

\subsection{Open questions and opportunities}
\subsubsection{Anatomical detail in musculoskeletal models}
\label{ch:open_questions_musc} 
It is unclear to what extent biological details of the musculoskeletal system are necessary to recapitulate the key features of neural dynamics.  In fact, much of the recent work building whole-body models of animals often used as model systems in neuroscience has focused on joint-based control and left incorporation of \textit{musculo}skeletal dynamics into models for future work \cite{2024aldarondoVirtualRodentPredicts, 2025vaxenburgWholebodyPhysicsSimulation}.  While these models have obtained impressive whole-body control, it is possible that incorporating musculoskeletal details will improve the ability of these models to capture properties of neural population dynamics.

\subsubsection{Biological fidelity in network architecture} 
As detailed in Sections 2, motor control emerges from macroscale circuits spanning multiple brain regions as well as the spinal cord. Hierarchical architectures are ubiquitous in vertebrate motor systems \cite{merel_hierarchical_2019}, and have proven to be an especially useful motif for developing flexible control in artificial embodied systems \cite{2024aldarondoVirtualRodentPredicts, 2025vaxenburgWholebodyPhysicsSimulation}.  Thus, embodied models that implement such circuits, will likely continue to be developed. Multi-regional RNNs trained in a goal-driven fashion have previously been used in motor \cite{2020michaelsGoaldrivenModularNeural} and cognitive \cite{kleinman2021mechanistic} settings. Such models during embodied control may elucidate the distributed mechanisms underlying action selection, movement invigoration, timing, planning, and sequencing. Moving forward, it will be exciting to see work emerging at the interface of practical limitations guiding modeling design choices and the potential for increasingly sophisticated cell-type-specific biological fidelity, especially in species with known connectomes.

\subsubsection{Testing hypotheses about biological sensorimotor control}
Physically simulated embodied control models trained with DRL offer a powerful platform to model key features of biological sensorimotor control, including sensorimotor delays and predictive mechanisms \cite{jiao_deep_2024, karashchuk_sensorimotor_2025}. Sensorimotor delays are ubiquitous in biological systems, but difficult to manipulate experimentally. They can, however, be incorporated naturally into simulation through control and physics timestep variation. Observation and action noise can also be directly manipulated to study their effects on learning and control \cite{osborne_sensory_2005, schumacher_dep-rl_2023}. Predictive mechanisms, evident even near the sensory periphery in muscle spindles and the retina \cite{dimitriou_human_2022, liu_predictive_2021}, are especially critical to mitigating delays during dynamic interactions with the environment, as in complex natural behaviors such as prey capture \cite{hein_algorithmic_2020, shaw_fast_2023}. Embodied DRL agents also allow exploration of prediction as an auxiliary objective, which has been shown to induce structured representations, and may ground behavioral observations in neuromechanical principles \cite{fang_predictive_2024}. These simulations uniquely offer full access to egocentric observation streams and motor outputs, providing a testbed for modeling the sensorimotor loop as a closed system of neural, musculoskeletal, and environmental interactions.

\subsubsection{Achieving flexible behavior}
Organisms can learn an impressive range of motor skills. Additionally, animals are able to quickly learn novel tasks that contain already-learned mechanisms. Understanding how the brain accomplishes this feat will require animals and model networks that perform multiple tasks such that the underlying representations can be observed. In cognitive settings, such models have been designed with compositional representations emerging~\cite{yang2019task, driscoll2024flexible}. However, models are lacking in the motor setting, with the exception of~\cite{merelDeepNeuroethologyVirtual2019b}. Experimentally, a shared manifold has been discovered in primate M1 while performing multiple motor tasks~\cite{gallego2018cortical}. Despite such progress, the underlying dynamical features such as possible compositional representations in biological and artificial networks performing multiple motor tasks is not well understood. Methods such as curriculum RL may be employed to further train embodied models on multiple tasks and observe network representations.

% Conclusion
\section{CONCLUSION}
Sensorimotor control has typically been analyzed from separate viewpoints - the study of the anatomy and physiology of distributed loops, neural population dynamics across regions, and the optimal control of bodies. Here, we summarize and synthesize these strands to better situate the emerging field of embodied control for understanding the neural basis of sensorimotor control. 

Going forward, progress will hinge upon models and experiments that span species, tasks, timescales, and brain regions. Equally pressing are questions about how much anatomical and biophysical detail is needed in body and network models, and how to use simulation-derived hypotheses to design incisive experiments. With computational tools and comprehensive datasets now mature, a unified, closed-loop study of embodied sensorimotor control is within reach.

% Summary Points
\begin{summary}[SUMMARY POINTS]
\begin{enumerate}
\item Vertebrate sensorimotor systems are implemented through hierarchical, parallel, nested, and distributed loops.
\item The brain likely performs computations through population level mechanisms, governed by non-linear dynamics on low-dimensional manifolds.
\item Optimal control models provide rational principles for understanding the internal processes and mechanisms of brain involved in sensorimotor processing. 
\item Recent developments in simulations of embodied control using deep reinforcement learning offer great promise for understanding biological sensorimotor control. 
\end{enumerate}
\end{summary}

% Future Issues
\begin{issues}[FUTURE ISSUES]
\begin{enumerate}
\item How important is incorporating anatomical detail in musculoskeletal models in order to understand the underlying neural substrate? Are there properties of neural dynamics that naturally emerge by recasting them in an optimal control framework for the control of bodies?
\item How can principles of nervous system design be applied to the design of network architectures in goal-driven frameworks?
\item In what ways can we probe biological systems with model-generated hypotheses for furthering our understanding of the neural basis of sensorimotor control?
\item How can embodied control systems be endowed with capacities for flexible control for multiple tasks in complex environments?
\end{enumerate}
\end{issues}

%Disclosure
\section*{DISCLOSURE STATEMENT}
The authors are not aware of any affiliations, memberships, funding, or financial holdings that
might be perceived as affecting the objectivity of this review. 

% Acknowledgements
\section*{ACKNOWLEDGMENTS}
We gratefully acknowledge funding from the National Institute of Health RF1DA056377.

% References
% \nocite{*}
\bibliographystyle{ar-style3}
\bibliography{refs}

\end{document}